\documentclass[letterpaper]{article} 

\usepackage{times}
\usepackage{helvet}
\usepackage{courier}
\usepackage{color}
\usepackage{graphicx}
\usepackage{booktabs}
\usepackage[vlined,ruled]{algorithm2e}
\usepackage{listings}
\usepackage{paralist,enumitem}
\usepackage{url}
\usepackage{fancyvrb} 
\usepackage{subcaption}

\newcommand{\example}[1]{\textit{#1}}

\newcommand{\citep}[1]{\cite{#1}}

\begin{document}

\title{Honey: A dataflow programming language for the processing, featurization and analysis of multivariate, asynchronous and non-uniformly sampled scalar symbolic time sequences}

\author{Mathieu Guillame-Bert\\ \\
        School of Computer Science\\
        Pittsburgh, United States\\
        mathieug@andrew.cmu.edu
        }

\maketitle

\begin{abstract}
We introduce HONEY; a new specialized programming language designed to facilitate the processing of multivariate, asynchronous and non-uniformly sampled symbolic and scalar time sequences. 
When compiled, a Honey program is transformed into a static process flow diagram, which is then executed by a virtual machine. 
Honey's most notable features are:
(1) Honey introduces a new, efficient and non-prone to error paradigm for defining recursive process flow diagrams from text input with the mindset of imperative programming. Honey's specialized, high level and concise syntax allows fast and easy writing, reading and maintenance of complex processing of large scalar symbolic time sequence datasets.
(2) Honey guarantees programs will be executed similarly on static or real-time streaming datasets.
(3) Honey's IDE includes an interactive visualization tool which allows for an interactive exploration of the intermediate and final outputs. This combination enables fast incremental prototyping, debugging, monitoring and maintenance of complex programs.
(4) In case of large datasets (larger than the available memory), Honey programs can be executed to process input greedily.
(5) The graphical structure of a compiled program provides several desirable properties, including distributed and/or paralleled execution, memory optimization, and program structure visualization.
(6) Honey contains a large library of both common and novel operators developed through various research projects.
An open source C++ implementation of Honey as well as the Honey IDE and the interactive data visualizer are publicly available.
\end{abstract}



\section{Introduction}
\label{sec:introduction}

Time series (TS; sequence of uniformly sampled numerical measurements) is a popular representation for storing and processing series of observations acquired from dynamical systems (e.g. weather forecasting, economic analysis, financial forecasting, quality control, signal processing, control, monitoring). TS are constructed by periodically recording the output of a sensor.
Time series from real world applications are often too large to be processed or analyzed manually by users, and each research and engineering community using time series have developed specialized algorithms, methods and software to process, analyze and interpret their time series.
While some methods have been found to be useful on multiple domains, most of time series processing remains largely done with specialized domain specific (and often expensive) tools with domain specific formats, embedded expert knowledge and domain specific interpretation frameworks.

Despite being popular, time series are not universally suited for the study of dynamical systems. For example, times series cannot represent directly time-point events, intervals, non-uniform sampling, or asynchronous multi-variate records. In this paper, we use a normalized extension of times series called Symbolic and Scalar Time Sequences (SSTS). SSTS allows the representation of multivariate, asynchronous and non-uniformly sampled symbolic and scalar records (see definition in section~\ref{sec:ssts}). 
As an example, an SSTS can represents multiple non-synchronized non-uniformly sampled times series with gaps, specific change points, time segmentation information, continuous recording and time-point events. All time series operators (e.g. moving average) can be extended, and sometimes augmented into the SSTS framework: For example, the \textit{moving average} of a time series $T$ is a new time series $T'$ with $T'_i$ defined as the mean of the last $n$ observations i.e. $T'_i = \sum_{j=i-n+1}^i T_j / n$. Instead, the moving average of an SSTS can be either defined as the average of the last $n$ observations (similarly as for time series), or as the average of all the observations in the tailing window of length $m$ temporal units.
Because of its freedom of expression, processing SSTS is often more complex than processing time series: While many operations on time series can be done with a single ``for loop'' or common matrix operations; the same operations on SSTS require more complex implementations, mostly because of the non-uniform sampling and non-synchronicity of the channels.

In this paper, we introduce a specialized programming language for the purpose of processing SSTS (and therefore time series). This language, called Honey, relies on a new paradigm designed to create process flow diagrams from imperative programming like input. Honey aims to provide a powerful tool for data analysts, researchers and engineers on the tasks of temporal data exploration, data cleaning, data feature extraction, retrospective data analysis, real-time monitoring, process prototyping and development. Honey's high level, specialized and concise syntax allows for efficient writing, reading and maintenance of complex programs. In addition, Honey is associated with a continuously growing library of operators developed over several years of work on various applied research projects. The library of operators includes various moving statistics (e.g. moving average, moving standard deviation, moving range), control charts, frequency analysis, filtering, correlation operators, temporal landmark testing, forecasting analysis, feature selections, feature reductions, pattern matching and machine learning algorithms.
Honey aims to be both easy to use and efficient to execute (execution time and memory consumption wise). Honey can be used in a standalone fashion, or it can be integrated into larger projects though its API. Finally, a Honey program can seamlessly be executed on a static dataset or a real-time stream of data. This feature allows Honey user to develop and test programs on static records, and then, to easily execute them directly on real-time applications.
Since its first version in 2011, Honey has been used to support research in wide variety of domains (including various medical domains, banking activity analysis, crowd behavioral analysis, human activity forecasting, social interaction analysis, vehicle maintenance monitoring, market trading, market monitoring, and various robotic projects) both for post-hoc and real-time analyses.

This manuscript contains an introduction to the Honey's paradigm, syntax and semantic. It also features four complete and illustrated real-world data analytic exercises solved with Honey.
This manuscript does not aim to be exhaustive, and many important aspects of the language are not covered, including batch processing, file formats, available operators, and interactive output visualization and exploration.
Instead, this document is complementary with the Honey documentation and tutorial available online~\cite{Honey}.

This paper is structured as follows:
Section~\ref{sec:ssts} describes the Symbolic and Scalar Time Sequences representation.
The section~\ref{sec:related_work} discusses related works.
Section~\ref{sec:hello_world} presents and explains a small ``Hello world'' Honey program. 
Section~\ref{sec:language} introduces the Honey programming language syntax and semantics.
Section~\ref{sec:examples} shows and discusses four examples of Honey programs. These examples aim to illustrate typical usage of Honey.
In section~\ref{sec:discussion}, we discuss the reason for Honey's development, the way it has been used over the last years, the reason of its current shape, and its relation to alternative analytic solutions.
Section~\ref{sec:conclusion} concludes this manuscript.

\section{Symbolic and Scalar Time Sequences}
\label{sec:ssts}

A Symbolic and Scalar Time Sequence (SSTS) is representation that extents Time Series. A SSTS is composed of a set of channels. Each channel is composed of a symbol (i.e. a name) and a set of records. A record is composed of a time-stamp and (optionally) a value. A time-stamp is a representation of time encoded as a double precision floating-point number. Honey does not make assumptions on the unit of the time representation (it can be days, years, seconds, microseconds, indexes, etc.). The value is a floating-point number representing a measurement (e.g. a temperature).

\begin{figure}
\centering
\includegraphics[scale=1]{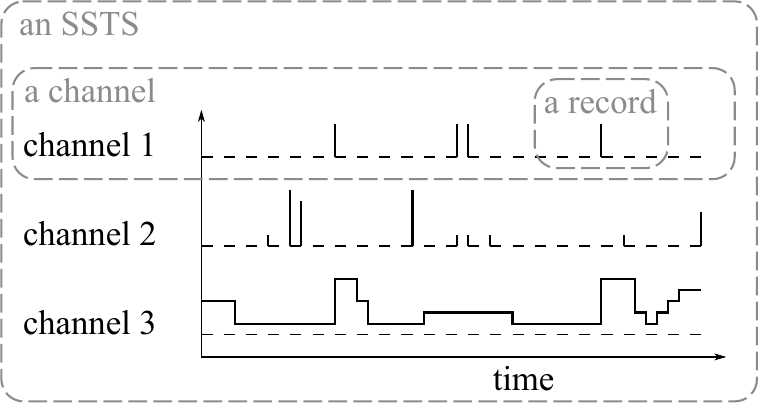}
\caption{Illustration of an SSTS.}
\label{fig:ssts}
\end{figure}

The figure~\ref{fig:ssts} shows an illustration of an SSTS.
In this figure, while being plotted differently, channels ``channel 1'' and ``channel 3'' are both SSTS channels (channel 1 is plotted with ``bars'', and channel 3 is plotted with ``stairs''). Honey does not make assumptions about the continuity of the sampling or the nature of the channels: A channel can be a continuous recording, a time segmentation, time-point events, etc.

\emph{Scalar channels} contain continuous (or repetitive) measurements.
By convention, the value of a scalar channel at time $t$ is the value of the latest record non-strictly anterior to $t$.
\emph{Event channels} contain event or change-point records.
SSTS represents scalar and event channels indistinctly, but the documentation of several Honey operators rely on this scalar/event conversion. A same channel can be treated as scalar by an operator and as an event by another one. By definition, a time series is a single scalar channel with uniform sampling.

Honey main input format is the .evt file. Each line of an .evt file defines a single record. Honey also supports .csv files where each line defines a record for all channels at a given time. This implies that .csv files can only define synchronized scalar channels (unless if using ``NA'' (non-available) values). Both .evt and .csv files can be streamed over telnet-like connections to and from Honey. Honey supports several other formats including an optimized binary version of .evt files.

\section{Related work}
\label{sec:related_work}

Programming languages are used to define computation processes (also called programs) while avoiding or abstracting low level hardware interaction (e.g. assembly, electronic design). Except for very specific problems, today, most program writing is done by the intermediate of a programming language. The popularity of programming languages can be explained by the increase of the power and capacities of computer hardware's, the increase of complexity of programs, and the improvement of compilers and programming languages.

Programming languages provide many desirable features for users, including paradigm abstractions, reduction of the propensity of errors, compact representation, inter-systems portability, code analysis and error checking. The majority of programming languages are Turing Complete, and all have the ability to solve the same class of problems. However, each programming language exhibits different features which make it adapted for more or less specific classes of problems. Therefore, the choice of a programming language is generally defined by the task to be solved as well as by the industry domain standard: For example, C is adapted for high performance low level system programming, R is good for small scale data analysis, Prolog is suited for combinatory exploration tasks, and Ada is designed for critical systems. The ``specialization'' of a programming language refers to the range of problems that can be solved easily. As an example, C, C++, Java, Python and FORTRAN are considered to be ``general'' programming languages, while Html+Css, Latex, Sql, UnrealScript, MT4, ChucK and Verilog are considered to be specialized programming languages. Specialized programming languages are generally significantly easier to use and more efficient to solve their target task than would be a more general programming language. In this paper, we are presenting a specialized programming language designed to process SSTSs and other equivalent temporal records. And while, any other Turing complete languages could be used instead, we demonstrate the efficiency and simplicity of use of our language for this task.

The following paragraph lists software solutions for processing and analyzing temporal databases: When handling high volume of data, users often use general high efficiency programming languages (e.g. C, C++, and FORTRAN), adapted architectures and hardware, and specialized libraries (also written with high efficiency programming languages). In the case of smaller problems and for fast prototyping, specialized numerical scientific programming languages are generally more suited (e.g. R, MatLab, SciLab, Octave, Mathematica and Maple). Many such languages have large active communities that develop and maintain specialized scientific libraries. Database programming languages (e.g. Sql, Clarion and DBase) are specialized programming language for querying and modifying databases. They are often used conjointly with another more general programming language.

While most programming languages are text based, various other representations have been explored. For example, visual programming relies on the drawing or arrangement of graphical elements to define programs. Graph programming refers to the family of graphical languages relying on the definition of a graph (i.e. a set of nodes connected by edges). Graph programming is generally considered easier to learn than text base programming language. However, graph programs can be slow to execute (in case of fine grain programming), hard to maintain on large programs, and specialized for certain tasks. Graph programming language is mainly divided into two families: Logic Flow and Data Flow programming languages. In a logic flow programming language, nodes represent states of the program, and edges represent the transitions between states. Logic flow programming languages are used for logic controller (Sequential function chart, Function block diagram), game development (Unreal Engine, Gamemaker) and general sequential control (DRAKON, Scratch and Flowgorithm). Instead, in a Data Flow programming languages, nodes represent individual processes (called agents), and edges represent the connection between the outputs of a process to the inputs of the next ones. Pieces of information called tokens travel trough the network and are processed by each node they encounter. Data Flow programming languages are popular in media processing (Pure Data, Max, 3ds Max Material, Blender Shading Composer), machine learning (Weka, Orange, RapidMiner), signal processing (System Studio, SPW, Simulink) and architecture definition (VHDL).

Data flow computation modeling refers to the study of hardwares or softwares implementation of Data Flow programming languages. Several models have been studied with different constraints, guarantees, speed, robustness and ease of use in the industry.
A popular version has been studied by Kahn~\cite{Kah74_dataflow}.
\cite{Moreira2014} and \cite{Najjar1999} provide introductions and a state-of-the-art review of the Data flow research. Data Flow models are particularly adapted for parallel or distributed computing: Each node (or agent) works independently from the rest of the network except from its parents and children (in some models). However, this property has a cost: Programs need to be adapted to flow processing, and the synchronicity between nodes can be a significant part of the computation work. Various text base programming languages have been studied to define data flows (Lustre~\citep{Halbwachs91_lustre}, Signal~\citep{LeGuernic91_signal}).

\section{``Hello World''}
\label{sec:hello_world}

Before presenting the Honey syntax and semantics, we show a simple example of a Honey program. This program takes as input a SSTS, and computes a 0.1 time unit tailing moving average (sometime called simple tailing moving average) independently on all the channels. The program outputs both the input records and the results of the moving average.
We suppose the input dataset to be provided in the file \textit{dataset.evt}, and the resulting SSTS is written in the file \textit{result.evt}.
The following listing shows the source code of the program.

\begin{figure}
\centering
\includegraphics[scale=0.8]{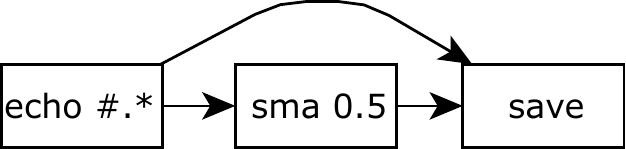}
\caption{Flow graph of the Hello World program.}
\label{fig:helloworld}
\end{figure}
\definecolor{Mycolor}{rgb}{0.3,0.3,0.3}
\lstset{
columns=fullflexible
, language=Perl
, numbers=left
, xleftmargin=1.7em
,basicstyle=
,frame=single 
,numberstyle=\small
,breaklines=true
,commentstyle=\itshape\color{Mycolor}
,basicstyle=\small
,showspaces=false
,deletekeywords=return
,aboveskip=0pt
,belowskip=0pt
,otherkeywords={include,group,calendar,count,test,multiTests,report_histIntersectEventState,skip,delay,derivative,range,sinceLast,report_amoc,filter,sample,active,sd,call,sma,global,endif,eq,passIf,recursivedelay,tma,ema,endfunction,function,layer,normalize,echoPast,rename,saveBufferedCsv,@data,save,segment,return,echo,recursive}
,postbreak=\raisebox{0ex}[0ex][0ex]{\ensuremath{\hookrightarrow\space}}
,tabsize=4
}
\begin{lstlisting}
@data input:"dataset.evt" output:"result.evt"
$all = echo #.*
$res = sma $all 0.1
$res += echo $all
save $res file:%output
\end{lstlisting}

Each line is explained in the following listing:
\begin{compactenum}
\item Specify the input and output of the program. Several input/outputs can be specified. The input and output of a program can also be specified outside of the program.
\item Set the \emph{channel variable} \$all to contain all the currently available channels (i.e. the input channels). Most Honey programs start with this statement.
\item Apply a simple tailing moving average independently on all the channels contained in the variable \$all. The resulting channels are put into the \$res variable. The name of the resulting channels are automatically generated: The names of the result on input channel ``channel1'' will be ``channel1\_sma[0.1]''.
Input channels are not assumed to be synchronized.
\item Append all the input channels to the \$res variable: The output dataset will also contain the original data.
\item Export all the channels from the variable \$res into the specified output file. This output file contains both the raw data and the result of the moving average.
\end{compactenum}

Figure~\ref{fig:helloworld} shows the process flow diagram obtained from this program after having been compiled.
In this example, the ``channel variable'' \$all and \$res behave similarly as variables in imperative programming languages (e.g. C, Python, Fortran). For example, the channel variable \$res refers to two different objects at lines 3 and 4.  \$res defines two different \emph{edges} between operations in the final process flow diagram. It is important to notice that while pipes in the final process flow diagram are not ordered (by definition), the order of the statements with channels variables is important: Reversing lines 3 and 4 in the listing would define a different program.

When a Honey program is executed on a dataset (in streaming execution mode -- see section~\ref{subsec:executionmodes}), operations are all executed in parallel, continuously processing inputs and generating outputs for the next ones. Each operator manages its own internal state and stored data. In our example, because the moving average operator requires keeping a record of all the elements in the current window, this ensures that when executed, the program keeps at most 0.1 time units of the each channel records' in memory at any time, independently of the size of the input dataset.
This example illustrates Honey's paradigm: Using imperative programming mindset to define process flow diagrams.

\begin{figure}
\centering
\includegraphics[scale=1]{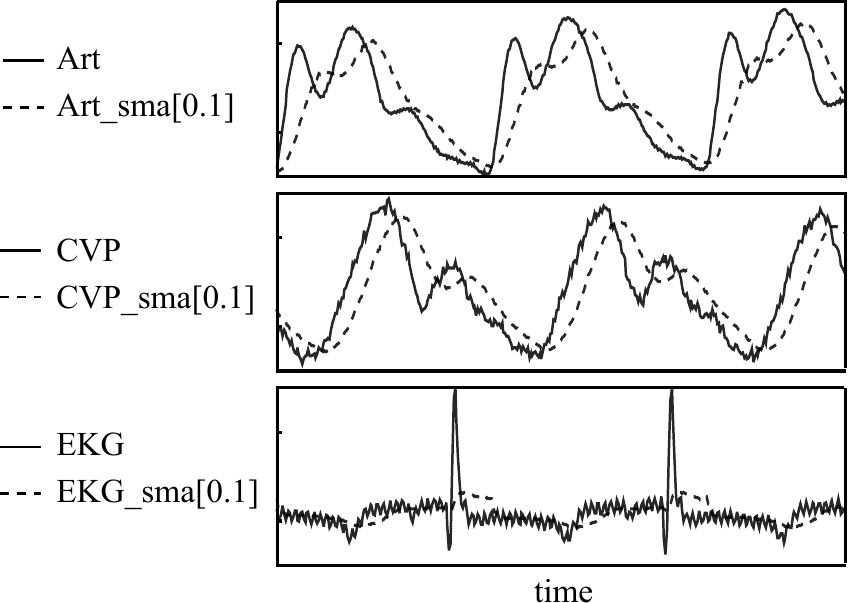} 
\caption{Output of the hello world program on a small medical dataset containing high frequency records of various vital signs. The input channels are drawn with solid plots. The resulting channels (simple moving average) are drawn with dashed plots.}
\label{fig:hw_plot}
\end{figure}

The figure~\ref{fig:hw_plot} shows the result of this program on a medical dataset. This dataset is a 1.5 second snap-shot of various physiologic vital sign (CVP, ECG and Arterial pressure) recorded at a frequency of 250Hz.

\section{The Honey language}
\label{sec:language}

In this section, we introduce the Honey programming language.
The next sub-section presents the syntax and semantic of the Honey programming language.
The reader is assumed to be familiar with imperative programming.
The following sub-sections presents specific aspects of the language (operators, equations, execution, user defined operator and recursions). 

\subsection{Syntax and semantics}
The source code of a Honey program is a plain text file. Each line of this file is one of the following items. For each item, we provide one or several examples.

\lstset{numbers=none}
\begin{compactitem}

\item A \emph{comment line} if the first non-white-space character of the line is \#. Comment lines are ignored by the compiler.
\begin{lstlisting}
# This is a commentary
\end{lstlisting}

\item A \emph{configuration statement} if the first character of the line is @. Configuration statements define how a program should be executed. The order and the location of the configuration statements has no importance in the source code. Configuration statements are optional and can be replaced with options on the Honey command line call when running a program.
\begin{lstlisting}
@data input:"dataset.evt" output:"result.evt
\end{lstlisting}

\item An \emph{operator statement} defines a single operator. A operator statement is composed of a operator name followed by arguments.
\begin{lstlisting}
sma $b 5 trigger:$c
\end{lstlisting}
Each operator requires specific arguments as defined in the documentation. Arguments can be \emph{anonymous} or \emph{named}, \emph{optional} or \emph{required}. In the previous example, the first two arguments of \example{sma} are two \emph{anonymous} and \emph{required} arguments, while the \example{trigger} argument is an \emph{optional} and \emph{named} argument.
Some operators do not require any arguments (e.g. a regular ticking generator operator) while some others do not have any outputs (e.g. an file exporter operator).
Optionally, a statement can specify a variable to receive the output (set of channels) with ``=''.
\begin{lstlisting}
$a = sma $b 5 trigger:$c
\end{lstlisting}
The output of an operator can also be merged with the content of an already existing variable with ``+=''.
\begin{lstlisting}
$a += sma $b 5 trigger:$c
\end{lstlisting}

\example{echo} is a special operator that simply repeats signals. \example{echo} is generally used to organize flows. When used with non-recursive signal variables, \example{echo} can be ``removed'': The two following lines are generally equivalent:
\begin{lstlisting}
$b += echo $a
$b += $a
\end{lstlisting}


\item The beginning or the end of a \emph{user defined operator/function}.
\begin{lstlisting}
function doSomething $a %b
...
endfunction
\end{lstlisting}

\item A call to a user defined operator.
\begin{lstlisting}
call doSomething $all 5
\end{lstlisting}

\item An inclusion of another Honey program.
\begin{lstlisting}
include "library.hny"
\end{lstlisting}

\item A definition of a \emph{non-channel variable}. A non-channel variable can contains a number or a string.
\begin{lstlisting}
set %a "Hello World"
set %b 1024
\end{lstlisting}

\item A ``if'' statement on non-channel condition. \emph{if} statements are evaluated during the compilation, and they are use to control the structure of the resulting flow diagram. 
\begin{lstlisting}
if =%i,0,$>$
...
endif
\end{lstlisting}
A \emph{if} statement cannot apply a condition on the contents of a channel. Instead, specialized operators should be used for any conditional processing of records according to their value or according to other channel values.

\item The beginning and the end of a \emph{group}.
Groups provides a visual enhancement when plotting the process flow diagram or when analyzing program execution. They have no impact on the result values.
\begin{lstlisting}
group "Computation of XYZ"
...
endgroup "Computation of XYZ"
\end{lstlisting}

\item The specification for a channel variable to be recursive (see section~\ref{subsec:recursion}). Recursive channel variables behave differently than non-recursive channel variables.
Recursive variables are powerful but prone to coding error. For this reason, variables are by default non-recursive.
\begin{lstlisting}
recursive $a
\end{lstlisting}

\item The specification for a variable inside of a user defined operator to be global. Global variables have a full scope over the entire program.
\begin{lstlisting}
global $a
\end{lstlisting}

\end{compactitem}
\lstset{numbers=left}

When compiled, a Honey program is converted into a process flow diagram. Operators in the Honey source code will define operators in the flow diagram, and channel variables in the Honey source code will define edges/pipes in the process flow diagram. A single channel variable can define multiple pipes, and a single operator in the source code can define multiple operators in the process flow diagram. For this reason, the source code of a Honey is often more compact and easier to understand that the resulting process flow diagram.

\subsection{Convention on operators}

Honey aims for programs to work similarly on static and real-time streaming datasets. For this reason, and for most of Honey operators, an operator's output value at time $t$ can only be derived from input records anterior to time $t$.
For example, the \example{sma} operator (simple moving average) computes a tailing moving average: The output at time $t$ is the average of the record values between times $t-w$ and $t$ (with $w$ the window length parameter of the \example{sma} operator).

In case of analysis were posterior data are necessary (e.g. retrospective analysis), users can use the special \example{echoPast} operator. This operator does not follow the tailing rule: EchoPast ``sends'' records into the past i.e. a records at time $t$ will be sent at time $t-w$ with $w$ the echoPast parameter. EchoPast can be used conjointly with other operators to defined non-tailing behaviors. The following example shows a centered moving average using the echoPast operator.
\begin{lstlisting}
@data input:"dataset.evt" output:"result.evt"
$all = echo #.*
$all = echoPast $all 5
$res = sma $all 10
save $res file:%output
\end{lstlisting}

Launching a Honey program with an echoPast operator on an real-time dataset will raise an error.

\subsection{Equations}

Honey supports three types of equations: Equations on records (e.g. the \example{eq} operator), equations on non-signal numerical values (starting with =), and equations on non-signal textual values (starting with \&).
Honey equation statement should be written in Reverse Polish notation (RPN) (e.g. ``=2,3,+,2,*'' = $(2+3) \times 2$ = $10$ ).

The following statement shows an example of equation on records. This statement takes the channels in \$a, and multiplies their values by 3 (i.e. if an input record at time $t$ has the value $v$, an output record will be create at time $t$ and with value $3 \times v$):
\begin{lstlisting}[numbers=none]
$b = eq $a "value,3,*"
\end{lstlisting}

The following statement is an example of equation on numerical non-channel variables. It computes a 10 time units moving average:
\begin{lstlisting}[numbers=none]
$b = sma $a =2,3,+,2,*
\end{lstlisting}

The following statement is an example of equation on string non-channel variables. It renames a channel to ``The little cat'':
\begin{lstlisting}[numbers=none]
$b = rename $a "&The ,little ,+,cat,+"
\end{lstlisting}

\subsection{Execution modes}
\label{subsec:executionmodes}

Honey supports three modes of executions:
\begin{compactitem}
\item The \emph{real-time execution mode} processes real-time stream of records from real-time input Honey connectors (e.g. telnet+evt, keyboard, calendar). Time is assumed to be the double precision floating-point Unix epoch time (i.e. the number of seconds since Thu, 01 Jan 1970 00:00:00 GMT). Input and intermediate records are kept in memory until the program knows they won't be used anymore. For example, if a program computes of a moving average with a window length of 10 seconds, only the records in the last 10s will be kept in the dedicated memory of the moving average operator.
\item The \emph{streaming execution mode} processes a static dataset (e.g. a dataset contained in a file on your hard-drive) in the same way as the real-time execution mode. The input dataset is parsed record by record (e.g. line by line in the case of .csv or .evt files). This mode generally ensures for the program execution to have a small memory footprint with the consequence that it is slower than the static section mode (see below).
\item The \emph{static execution mode} computes the results of each operator, one at a time in a topological order. Because operator's computation can often  be significantly optimized when all the data is available at once, and because the communication between operators can be a significant overhead, this mode is significantly faster than the streaming execution mode. However, since entire datasets (or segments of datasets in case of batched processing) need to be loaded in memory, this mode requires more memory than the streaming execution mode.

In our current Honey implementation, static execution mode does not support recursive programs because a recursive loop in a data flow program requires all the operators of the loop to be executed simultaneously. However, future Honey version will be able to execute recursive programs in hybrid static execution mode by processing individual recursive components in streaming mode while processing the overage program in static execution mode.

In practice, since the process flow diagram is static (i.e. a process flow diagram does not change during the execution), Honey can improve the global memory consumption of a program: Operator results are released from memory as soon as Honey detects they will not be used anymore by any other operators. In case of batching execution (processing of several independent datasets), the static mode will ensure that only one batch of data is loaded in memory at any time.
\end{compactitem}

The next two paragraphs illustrate the step by step execution of a program in streaming and static execution modes.

\begin{lstlisting}
@data input:"dataset.csv" output:"result.evt"
$a = echo #.*
$b = tma $a 2
$c = sma $b 2
$a += ema $b 2
save $a file:%output
\end{lstlisting}

The content of \textit{dataset.csv} is shown below. This dataset contains a single channel ``toto'' with three records at times 1, 2 and 5.

{
\small
\begin{Verbatim}[frame=single]
time;toto
1;1.0
2;1.1
5;1.2
\end{Verbatim}
}

\subparagraph{Static mode}
\begin{compactenum}
\item Load the data from \textit{dataset.csv} into memory.
\item Apply the tma operator for all the records for each of the input channels.
\item Apply the sma operator for all the records generated by tma.
\item Release the results of tma from memory.
\item Apply the ema operator for all the records generated by sma.
\item Release the results of sma from memory.
\item Save the result as well as the input dataset to the output file.
\item Release the results of ema from memory.
\end{compactenum}

\subparagraph{Streaming mode}
\begin{compactenum}
\item Read the first record (toto,1,1.0) from the file \textit{dataset.csv}, and ``send'' it to the operator tma. The tma operator keeps a copy of this record.
Send the input record to the save operator.
The save operator writes this record in the output file.
Send the output record of tma to sma. The sma operator keeps a copy of this record.
Send the output record of sma to ema. The ema operator keeps a copy of this record.
Send the output record of ema to save. The save operator writes the record to the output file. 
\item Read the second record (toto,2,1.1), and send it to the operator tma.
Send the record to the save operator. The tma operator keeps a copy of this record.
Send the output record of tma to sma. The sma operator keeps a copy of this record.
Send the output record of sma to ema. The ema operator keeps a copy of this record.
Send the output record of ema to save. The save operator writes the record to the output file.
\item Read the third record (toto,5,1.2), and send it to the operator tma.
Send the record to the save operator. The tma operator keeps a copy of this record.
The tma operator releases the two first records because they are out of the current time window [3,5].
Send the output record of tma to sma. The sma operator keeps a copy of this record.
The sma operator releases the two first records.
Send the output record of sma to ema. The ema operator keeps a copy of this record.
The ema operator releases the two first records.
Send the output record of ema to save. The save writes the record to the output file.
\end{compactenum}

\subsection{User defined operators}

As in most programming languages, users can define new operators either with the \example{function} keyword, or by extending a new C++ class in the Honey implementation. This sub-section presents the first solution.

The definition of a new operator is done with the keywords \example{function} and \example{endfunction}. When compiling a program, new operators' code is unrolled into the flow diagram. For example, if a user defined function is called twice, the flow diagram corresponding to the function will be repeated twice in the final process flow diagram. The memory footprint of this solution is often negligible, and ensures the result of the compilation to be a flat (i.e. non hierarchical) flow diagram.

User can define the result of an operator with the \example{return} keyword. Unlike most other programming languages, \example{return} does not stop the ``execution'' of an operator, and an operator can contain several active return statements.

\begin{figure}
\centering
\includegraphics[scale=0.8]{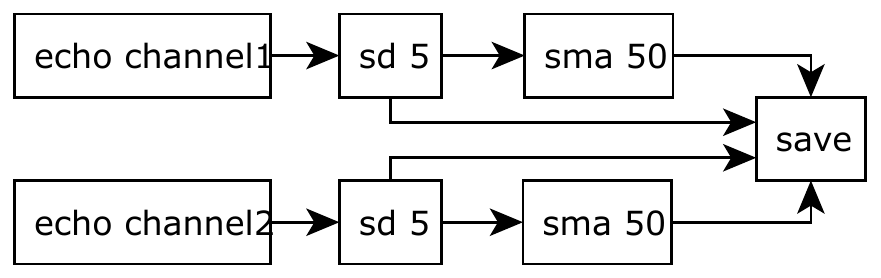}
\caption{Flow diagram of the function example.}
\label{fig:function}
\end{figure}

The following example shows the definition of a user defined operator as well as the resulting flow diagram shown in figure~\ref{fig:function}. In this listing, the work of functions f and f\_bis are exactly equivalent (even if f\_bis contains two \example{return} statements). However, since f\_bis is never called, it does not appear in the final flow diagram.

\begin{lstlisting}
function f $a %w
	$r = sd $a %w
	$r += sma $r =%w,10,*
	return $r
endfunction

function f_bis $a %w
	$r = sd $a %w
	return $r
	$r = sma $r =%w,10,*
	return $r
endfunction

$c1 = echo "channel1"
$c2 = echo "channel2"
$res = call f $c1 5
$res += call f $c2 5
save $res file:"result.evt"
\end{lstlisting}

\subsection{Recursion}
\label{subsec:recursion}

Honey supports two types of recursion: \emph{operator recursion} and \emph{record recursion}.

\emph{Operator recursion} happens when a user defined operator call itself recursively (directly or indirectly). As explained in the previous section, the flow diagram of the user defined operator is unrolled during compilation, and infinite recursions can be detected at compilation time. The final flow diagram does not contain cycles.

\emph{Record recursion} is a directed cycle in the final flow diagram. Through this cycle, records are re-injected ``upstream'' in the flow diagram. Unlike operator recursion, record recursion does not involve unrolling of an operator, and infinite recursion cannot be detected during the compilation. Honey syntax aims to protect the user against unwanted and erroneous recursions, and record recursion is only done with variables specified as recursive by the user (with the \example{recursive} keyword).

\begin{figure}
    \centering
    \begin{subfigure}[b]{\columnwidth}
        \includegraphics[scale=0.8]{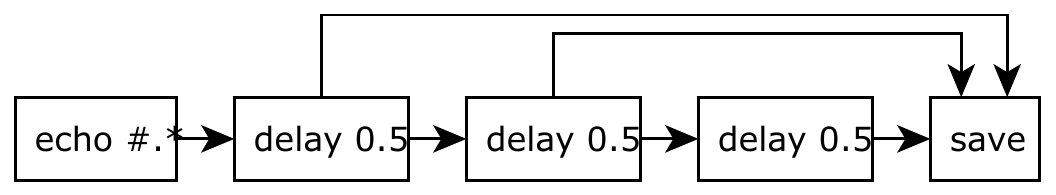}
        \caption{Operator recursion}
        \label{fig:gull}
    \end{subfigure}
    \begin{subfigure}[b]{\columnwidth}
        \includegraphics[scale=0.8]{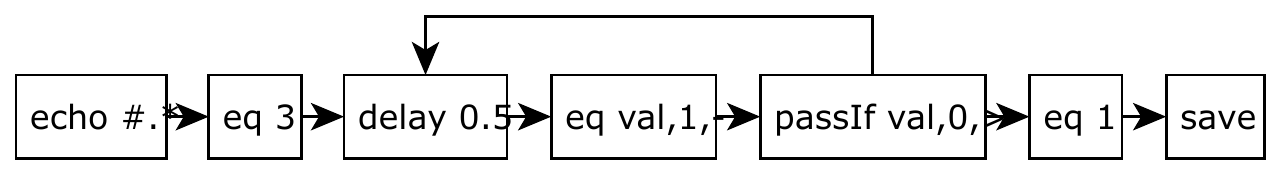}
       \caption{Record recursion}
        \label{fig:tiger}
    \end{subfigure}
    \caption{Flow diagrams for the operator recursive and record recursive solutions.}\label{fig:recursions}
\end{figure}

The following example shows two programs that repeat three times all input event records with an interval of 0.5 time units. The first program solves this exercise with operator recursion. The second program solves this exercise with record recursion. The flow diagrams of both programs are shown in figure~\ref{fig:recursions}.

Operator recursion solution:
\begin{lstlisting}
$a = echo #.*
function f $b %i
	global $result
	$b = delay $b 0.5
	$result += $b
	set %i =%i,1,-
	if =%i,0,>
		call f $b %i
	endif
endfunction
call f $a 3
save $result file:"result.evt"
\end{lstlisting}

Record recursion solution:
\begin{lstlisting}
$a = echo #.*
recursive $x
$a = eq $a 3
$x += $a
$b = delay $x 0.5
$b = eq $b "value,1,-"
$b = passIf $b "value,0,>="
$x += $b
$result += echo $b
save $result file:"result.evt"
\end{lstlisting}

\section{Examples of Honey programs}
\label{sec:examples}

In this section, we present four non-trivial tasks (i.e. tasks that cannot be solved by a single Honey operator), and their solutions using Honey.
All these tasks come from real problems and data analyses.
For each task, a dataset is presented, a task is described, a solution Honey program is detailed, and the results are plotted.
These examples aim to show the power of the SSTS representation and the simplicity and power of Honey to solve complex real world problems.
The datasets, the Honey programs and the interactive plots of each task are available online~\cite{Honey} in the section ``Honey paper examples''. It is recommended for the reader to run and experiment with these examples.
The tasks are:

\begin{description}
\item[Feature generation:]
Given a non-synchronized and non-continuous record of vital signs of a patient in a hospital bed (i.e. heart rate, respiratory rate, oxygenation level) sampled approximately every 20s, and a set of timestamped physician annotations of cardio-respiratory health instability (e.g. abnormally high or low heart rate).
a) Compute various tailing moving statistical features with different window lengths.
b) Study the individual predictive power of each feature to the task of forecasting the physician labeled instabilities periods.
c) Align the channels to the heart rate channel, and export the results into a .csv file.

\item[Heartbeat segmentation:]
Given a record containing raw continuous EKG and Central Venus Pressure (CVP) records sampled at 250Hz.
The CVP is the instantaneous blood pressure at the entry of the heart.
a) Detect heartbeats.
b) Compute heart rate.
c) In a separate .csv file for each intra-heartbeat interval, extract and export a snapshot of the raw channel readings.
d) Compute the average CVP for each intra-heartbeat interval, and export it to a .csv file.

\item[Abnormality detection and calendar correlation:]
Given a three year record of several proximity sensor readings located in an office.
a) Study the distribution of the activity with the hour of the day and day of the week. 
b) Build an incremental model of normality of the activities and use this model to detect abnormalities.
c) Study the distribution of abnormalities according to the hour of the day and day of the week.

\item[Pattern recognition:]
Given a record containing raw continuous CVP, EKG and airway pressure sampled at 250Hz.
a) Look for up-down-up-down patterns in a CVP record. This pattern is equivalent to the ``ACV'' pattern convention used to study CVP waveform in the medical literature.
b) Detect beginning of expiration from the airway pressure (pressure in the lungs).
c) In a .csv file, one row for each heartbeat, export the increase of atrial pressure (computed from the CVP tracking) and the duration since the last beginning expiration (computed from the respiration tracking). Use R to create a 2d squatter plot of these two components.
\end{description}

\subsection{Feature generation}
As input, we consider a file \textit{vital.evt} containing several days of various vital sign readings (heart rate, respiratory rate, and oxygenation level) from a patient on a hospital bed. Vital signs are recorded independently approximately every 20s. Channels have ``gaps'' (periods without records, e.g. when the patient went to the bathroom). A second file \textit{annotation.evt} contains annotations of the beginning of instability periods that we want to understand.

The task is to compute the moving average and moving standard deviation of each channel with windows of length of 5min and 1h, and to evaluate the forecasting power of each of the new computed features on and the instability annotations using AMOC (Activity Monitor Operator Characteristic) and a T-ROC (Temporal Receiver Operating Characteristic) plots.
Additionally, we want to synchronize the three channels according to the heart rate records, compute new channels indicating (continuously) the presence or absence of a reading in the input channels in the previous 5 minutes (also called activity detection), and export the synchronized result into a .csv file. 
The program solving this task is given below. Line wrapping is represented by the $\hookrightarrow$ symbol.

\begin{lstlisting}
# Specify the two input files.
@data input:"vital.evt;annotation.evt"
# Select all available input channels.
$all = echo #.*
# Selects channels RR, HR and SPO2.
$vitals = filter $all "(RR|HR|SPO2)"
# Selects channel event.real_alert.
$annot = filter $all "event.real_alert"
# Define two numerical variables containing the number of seconds in 5min and 1h.
set %5Minutes =60,5,*
set %1Hour =60,60,*
# Compute moving averages and moving standard deviations in the last 5min and 1h.
$feat = sma $vitals %5Minutes
$feat += sma $vitals %1Hour
$feat += sd $vitals %5Minutes
$feat += sd $vitals %1Hour
# Compute AMOC and T-ROC between the computed features and the annotations. Save the results (plots, reports, statistical test results, etc.) in the given directory.
report_amoc trigger:$feat target:$annot file:"results/amoc"
# Select the heart rate channel.
$HR = filter $vitals "HR"
# Sample the input channels at the time of the heart rate records.
$export = sample $vitals trigger:$HR
# Compute the activity of each channel.
$active = active $vitals %5Minutes margin:%5Minutes
$export += sample $active trigger:$HR
# Export the sampled channels into a .csv file.
saveBufferedCsv $export file:"results/data.csv"
\end{lstlisting}

\begin{figure}
\centering
\includegraphics[scale=0.8]{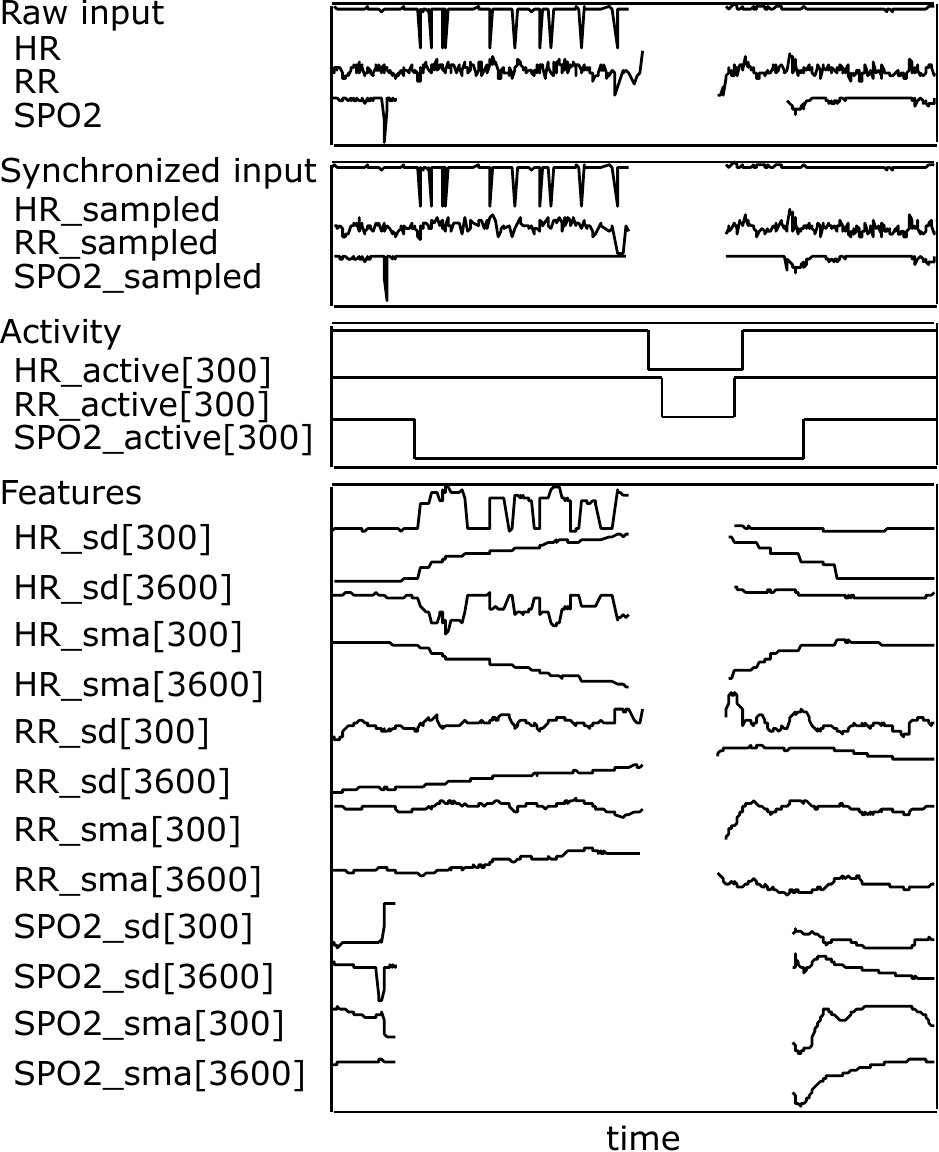}
\caption{Results from the task 1.}
\label{fig:task1}
\end{figure}

The figure~\ref{fig:task1} shows a small sample of the raw data, computed features and synchronized vital signs. This figure has been computed using Honey environment. The configuration of this plot is available at \cite{Honey}.

\subsection{Heartbeat segmentation}

As input, we consider a dataset containing continuous high frequency (250Hz) reading of CVP and EKG. The task is to detect heartbeats from the EKG readings, compute the heart rate, export a snapshot of the raw input for each heartbeat interval, and compute the average CVP heartbeat by heartbeat. The program solving this task is given below.

\begin{lstlisting}
@data input:"vital.csv"
$vitals = echo #.*
# Selects the channels 
$ekg = filter $vitals "EKG"
$cvp = filter $vitals "CVP"
# = Detect heartbeats
# Range of the ekg in the last 0.05s
$a = range $ekg 0.05
# Normalization from the mean and standard deviation estimated in the last 2s.
$b = normalize $a 2 type:meansd
# Generate an event when the value of the channel from $b crosses-up the value 2.
$c = layer $b thresholds:2 output:up
# Rename the channel to "heartbeat"
$heartbeat = rename $c "heartbeat"
# = Compute heart rate
# Compute the time interval since last record
$a = sinceLast $heartbeat 5
# Convert the interval into frequency by minutes.
$a = eq $a "60,value,/"
# Rename the channel to "heartrate"
$a = rename $a "heartrate"
# Remove records with values smaller than 1 or larger than 180
$heartrate = passIfFast $a minValue:1 maxValue:180
# Export the raw records of each heartbeat in a separate file.
saveBufferedCsv $vitals file:"snapshots/sh_<index>.csv" trigger:$heartbeat
# Save heartbeats events and heat rates channels.
$tosave = $heartbeat
$tosave += $heartrate
save $tosave file:"result.evt"
\end{lstlisting}

\begin{figure}
\centering
\includegraphics[scale=0.8]{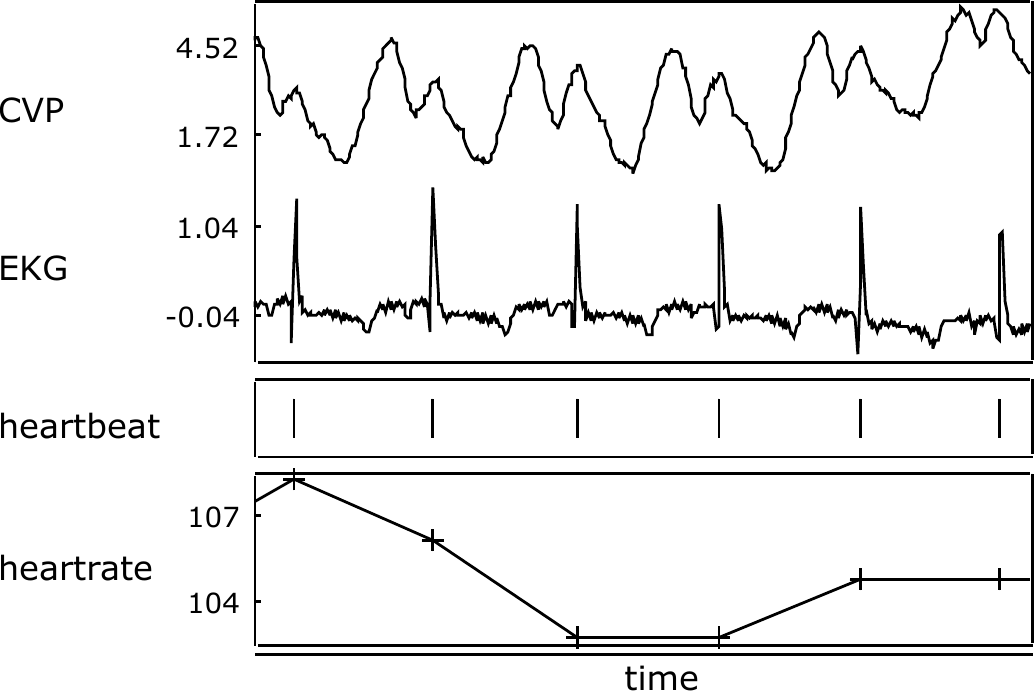}
\caption{Results from the task 2.}
\label{fig:task2}
\end{figure}

The figure~\ref{fig:task2} shows a small sample of the raw data, detected heartbeats and computed heart rate.

\subsection{Abnormality detection and calendar correlation}
As input, we consider an .evt file containing activation of proximity sensors installed in the hallway of an office building. A single record is generated whenever a person passes near a sensor. Each sensor is attached to one channel. This dataset is called the Merl dataset (Mitsubishi Electric Research Labs dataset) and it is available online~\cite{Merl}. The first task is, every 10 minutes, to compute the 1 hour tailing moving count of activity of each sensor. Next, every 10 minutes, we estimate a student's T distribution from the anterior moving counts conditioned on the day of the week and time of the day. The current moving count of each sensor is compared to its model and a set of p-values are estimated. Such p-values can be interpreted as abnormality detection scores -- a low value is indicative of an abnormal observation. The p-values from all sensors are aggregated using the Holm Bonferroni correction. The resulting value is the normalized number of rejected p-values. This number can be interpreted as the global abnormality of activity in the building. A threshold is used to define a boolean abnormality status. The daily and hourly distribution of abnormalities is finally studied. The program solving this task is the following:

\begin{lstlisting}
@data input:"sensor.bin" output:"abnormality.bin"
# Load the sensor readings
$all = echo #.*
# Select six of the sensors
$sensors = filter $all "s(386|330|406|265|375)"
# Generate a record every 10 minutes.
$every10mn = tick 600
# Compute a 1h moving count for each sensor.
$mcount = count $sensors 3600 trigger:$every10mn
# Generating calendar change points and status i.e. generate specific events for change of hour, day and month.
$cal = calendar produce:days,hours,months
# Select the event channel specifying the beginning of a new week.
$newWeek = filter $cal "event\.day_is_Monday"
# Every 10 minutes and for each sensor, estimate a student distribution parameters from past observations with same relative time difference to the beginning of the week. Compute and return the -log p-value between current observations and the estimated distribution. For example, an observation at 8:30am on Tuesday will be compared to all past observations at 8:30am on past Tuesdays. The operator also computes the confidence bounds.
$tests += test $mcount landmark:$newWeek trigger:$every10mn method:STUDENT
# Only keep the minus log p-values.
$indiv_abnormalities = filter $tests ".*mlogPValue"
# Aggregate all the -log p-values of abnormality using the Holm Bonferroni correction.
$abnormality = multiTests $indiv_abnormalities correction:HOLM_BONFERRONI alpha:0.01
# Threshold for abnormality alerts
set %threshold 0.2
$abnormality_t = passIfFast $abnormality minValue:%threshold
# Only keep one abnormality every 1h
$abnormality_t = skip $abnormality_t 3600
# Estimate the distribution of abnormality according to the hour of the day.
$cal_hour = filter $cal "state\.hour_is_.*"
report_histIntersectEventState event:$abnormality_t state:$cal_hour file:"report_hour.txt" 
# Estimate the distribution of abnormality according to the day of the week.
$cal_day = filter $cal "state\.day_is_.*"
report_histIntersectEventState event:$abnormality_t state:$cal_day file:"report_day.txt" 
# Save the results
$result = $cal
$result += $mcount
$result += $indiv_abnormalities
$result += $abnormality
saveBufferedBin $result file:%output
\end{lstlisting}

\begin{figure}
\centering
\includegraphics[scale=0.8]{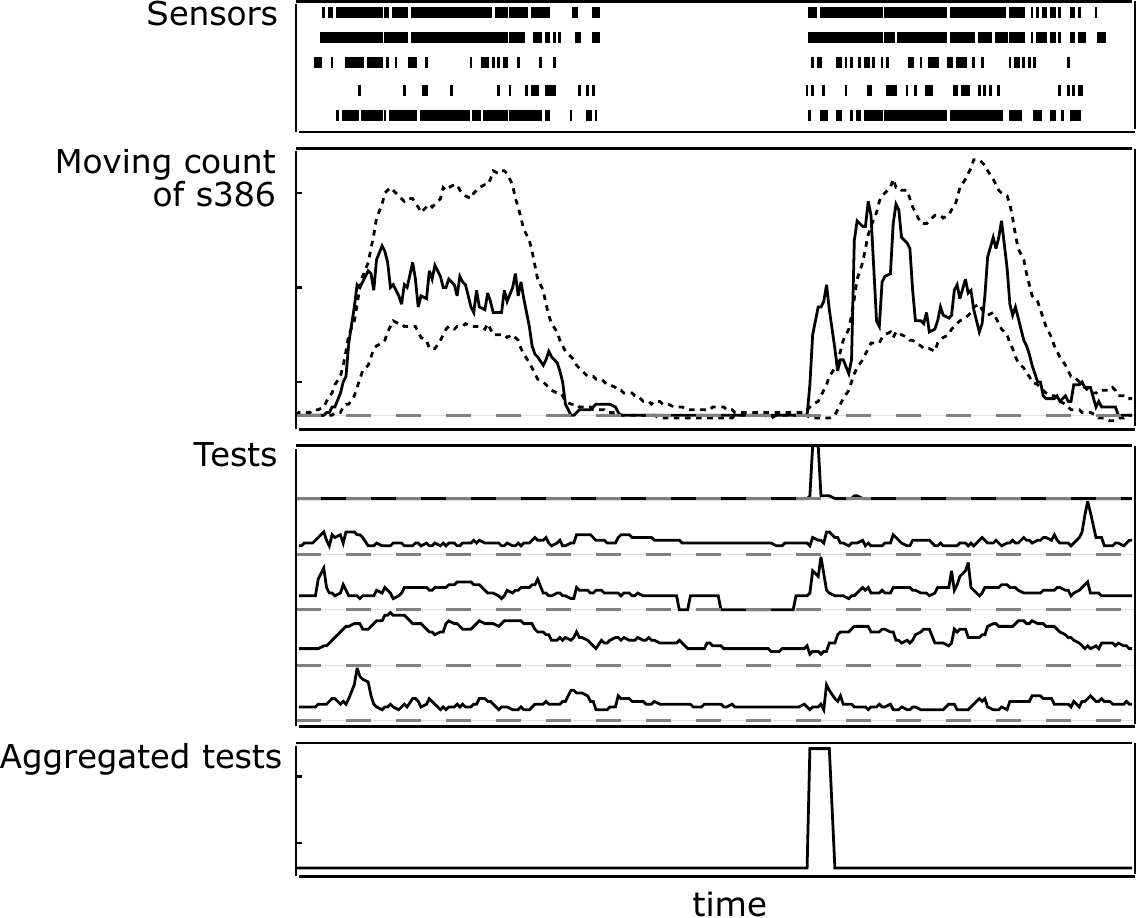}
\caption{Snapshot of sensor readings, moving counts, tests, confidence bounds and aggregated tests from the task 3.}
\label{fig:task3}
\end{figure}

The figure~\ref{fig:task3} shows the sensor readings, moving counts, tests, confidence bounds and aggregated tests.
The table~\ref{tab:task3} shows the distribution of abnormalities according to the day of the week (i.e. the file \textit{report\_day.txt}).

\begin{table}
\centering
\caption{Distribution of abnormalities according to the day of the week (\textit{report\_day.txt}).}
\label{tab:task3}
\small
\begin{lstlisting}
Number of types of events : 1
Number of types of states : 7
Event "multitest_0.01_HBON_0.01_skip_last[3600]"
Number of event instances: 2253
state.day_is_Sunday : 409 (18%)
state.day_is_Monday : 305 (13%)
state.day_is_Tuesday : 303 (13%)
state.day_is_Wednesday : 306 (13%)
state.day_is_Thursday : 318 (14%)
state.day_is_Friday : 305 (13%)
state.day_is_Saturday : 307 (13%)
\end{lstlisting}
\end{table}

\subsection{Pattern recognition}

As input, we consider a dataset containing continuous high frequency (250Hz) measurements of CVP, EKG and arterial pressure vital signs. The task is to track the CVP ``ACV'' patterns in the CVP waveform, to track the respiration cycle from the airway pressure waveform, and to create two dimensional scatter plot that shows, for each heartbeat, the relation between the Atrial pressure increase and the time since the last beginning of expiration. This comparison is important because during the breathing cycle, the change of pressure in the lungs impacts the blood pressure in the heart. The program solving this task is the following:

\begin{lstlisting}
@data input:"vital.csv"
$vitals = echo #.*
# Selects the channels 
$ekg = filter $vitals "EKG"
$cvp = filter $vitals "CVP"
$airway = filter $vitals "Airway pressure"
# User defined operator to detect heartbeats (same as task 2)
function computeHB $ekg
	$a = range $ekg 0.05
	$b = normalize $a 2 type:meansd
	$c = layer $b thresholds:2 output:up
	$hb = rename $c "heartbeat"
	return $hb
endfunction
# Detect heartbeats.
$hb = call computeHB $ekg
# Track the CVP waveform for each heartbeat: Track up-down-up-down patterns. The atrial pressure increase is the amplitude of the first up-down peak.
$d_hb = delay $hb 0.05
$cvp_seg = segment $cvp bound:$d_hb pattern:"U:X,D:A,U:Y,D:V"
# Select the increase of amplitude to A (Atrial).
$AV_r = filter $cvp_seg ".*_Dv_A"
# Track begening of expiration
$a = tma $airway 0.1
$b = derivative $a
$c = layer $b thresholds:0 output:down
$d = passIf $c arg1,7,> arg1:$a 
$begin_expir = rename $d "begin_expiration"
# For each heartbeat, export the Atrial pressure increase and time since expiration.
$tosave = $AV_r
$tosave += sinceLast $begin_expir 10 trigger:$AV_r
saveBufferedCsv $tosave file:"result.csv"
\end{lstlisting}

\begin{figure}
\centering
\includegraphics[scale=0.5]{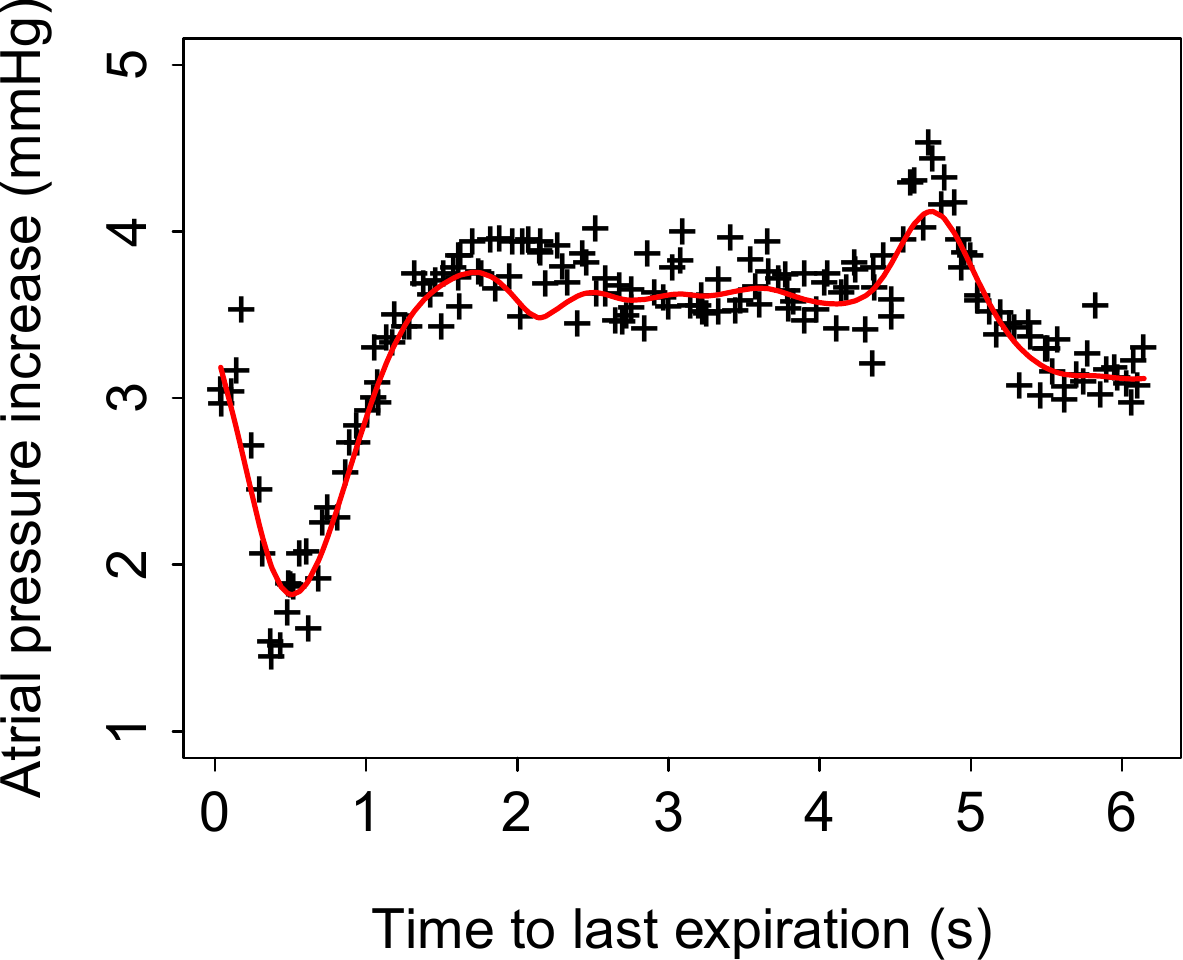}
\caption{Time since beginning of the last expiration and increase of atrial pressure of each heartbeat. The interpolation is based on a spline.}
\label{fig:task4resp}
\end{figure}

Figure~\ref{fig:task2} shows a snapshot of CVP, EKG and airway pressure, CVP tracking, respiration tracking, time since start of expiration and increase of atrial pressure. Figure~\ref{fig:task4resp} shows for each heartbeat, the relation between time since start of expiration and increase of atrial pressure.

\begin{figure}
\centering
\includegraphics[scale=0.8]{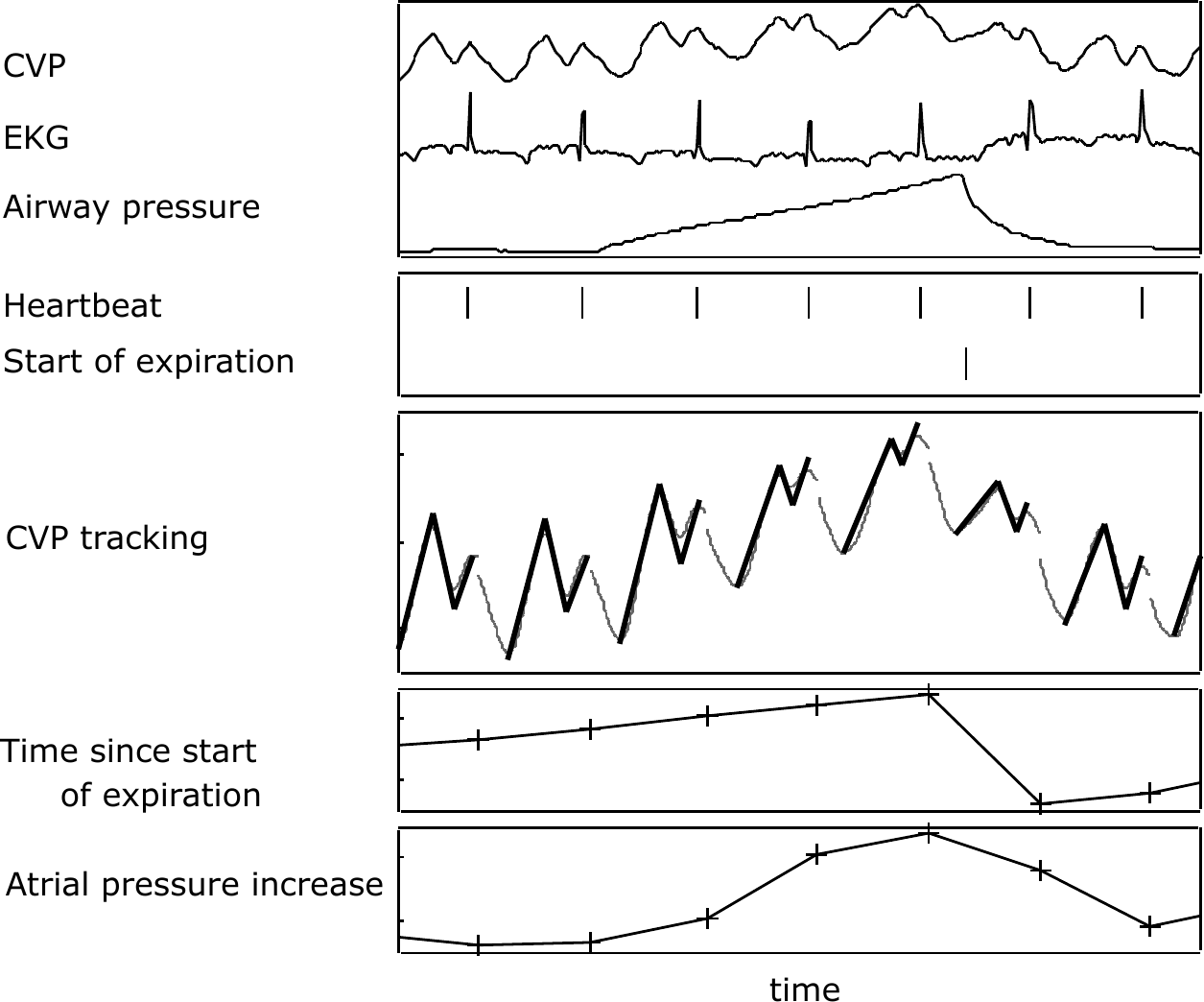}
\caption{Snapshot of CVP, EKG and airway pressure, CVP tracking, respiration tracking, time since start of expiration and increase of atrial pressure.}
\label{fig:task4}
\end{figure}

\section{Discussion}
\label{sec:discussion}


\subsubsection*{History of Honey}
The first version of Honey (initially called Event Processor) was created to handle seemingly real-time streams and static records of Forex (Foreign exchange market) pair values. Honey programs were initially written in xml, each operator was executed in a separate operating system process, and only a small number of operators were available. Honey outputs were primarily used as the input of the Titarl algorithm \cite{Titarl} to solve forecasting problems. Honey was next used to work with ``smart houses'' datasets (records of human activity in a house equipped with sensors). Next, Honey was use to process records of human robot interaction. Honey was next used to process vital sign records of patient in hospitals. During these various exercises, it appeared that many operators could be re-used across several domains. In addition, the direct availability of these operators, the ease with which they could be combined, and the structure of the framework made such experimentation fast (and flexible). Since that time, Honey has been improved through successive project applications, new operators were created, the syntax and semantics were improved, and the IDE environment was developed and refined.

\subsubsection*{Why Honey is simply not a library?}
Readers might wonder why Honey was designed as a programming language instead of a simple library. The reason is simple: To be executed a program in real-time or in static streaming execution mode, the Honey virtual machine needs to be fully aware of the entire flow diagram at the beginning of the execution. Therefore, a library that would allow calling each operator sequentially would not allow real-time or in static streaming execution mode. At best, the library would require specifying the entire process before the beginning of the execution, but this solution would be equivalent and less practical than writing a program/configuration file. In addition, having to ``hard code'' the process definition would impeach its flexibility, and require recompilation the calling program every-time the process definition was changed.
Instead, knowing the entire processing chain allows Honey to optimize the memory footprint of the execution. And, having an integrated standalone environment contributes to having a quick-to-use tool.

\subsubsection*{API and integration}
However, Honey can still be used as a library from inside of another program through its API. Through the API, a user can load, compile and execute Honey programs. This feature is especially useful for tight integration and for avoid easily to duplicate records in memory. Such collaboration between two languages is generally used for large projects that require flexible user inputs, and it has been found to be a powerful solution for fast prototyping and experimentation. The most famous example of such osmosis is certainly the Lua programming language~\cite{Lua} which is used in many industrial applications and many commercialized video games (more than $50\%$ of published video games in 2009).

\subsubsection*{Temporal and non-temporal hybrid analyses}
In many exercises, inputs are represented as large temporal datasets (e.g. time series), the first layers of the analysis require the processing of a large amount of data in the temporal domain, while the last layers are non-temporal and require the processing much smaller amount of data. Honey is routinely used in such exercises to quickly do the work of the first temporal layers of the analysis. Honey's results are then exported to a general numerical analysis environment were the remaining of the analysis is done (e.g. R, Matlab, Scilab, Pandas).

\section{Conclusion}
\label{sec:conclusion}

This paper presents Honey; a data flow programming language for the processing of multivariate, asynchronous and non-uniformly sampled scalar and symbolic time sequences. It was developed both for specific research and out of a desire for normalization and re-usability of data analytic workflows. Honey has been designed with a number of novel features which makes it a unique and powerful data analytic tool for its area of specialty. Honey novel features are:

\begin{compactenum}
\item A compact and specialized syntax for fast and efficient code writing, reading and maintenance. As an illustration, one of our past analyses involved a problem similar to task 4. The task was initially solved with a 872 lines Python program. As it appeared, because of the exploratory nature of the exercise, because the code was modified continuously, and because our experimentation did not require production quality code, the program became hard to maintain, understand, modify and experiment with. Additionally, the program was also slow. Rewritting it in C would have certainly made it faster but also more complex and less flexible. Instead, the program was re-written using a forty lines in Honey. This conversion solved many of the existing bugs, it required a very small amount of time to write and test, and it allowed us to quickly perform many new experiments in a very short time.

\item Honey programs are guaranteed to run similarly on static and read-time streaming datasets. This feature protects the developers from bugs that would only appear if and when an experimental code designed on static data was converted for work on online streaming data. Additionally, data analysts can more easily try their work on ``real world scenarios''.

\item Honey includes an IDE that contains a data visualization and an exploration environment. While writing a Honey program, the intermediate and final results can be directly and interactively explored. This feature allows fast design, debugging and experimentation. It also provides a solution to the problem of ``blind analytics'': Data analysts building invalid analysis due to incorrect assumptions on the data. Instead, and using Honey, users get a better understanding of the dataset without extra effort. They can also collaborate more efficiently with colleagues and domain experts. Such ``aware'' development often leads to better designed analysis, and reduces the risk of hidden errors.

\item Honey is efficient. Honey is implemented in C++ which make it significantly faster than interpreted (Matlab, R, etc.) or virtualized languages (Java, Python).
While most scientific environment languages (Matlab, R, etc.) have high efficient C implementations of algebraic functions, the high branching nature of operators required to process non-synchronous and non-uniformly sampled records only marginally benefits from these implementations. In addition, unlike scientific environments, Honey program execution optimizes the memory footprint. Comparisons have shown large gain both in memory and speed. Finally, Honey can be executed in seamlessly in greedy execution mode for datasets larger that the available memory.

\item Once compiled, a Honey program is converted into a process flow diagram. This graphical structure can be used to visualize and understand a program.

\item Finally, over the last seven years, we made a significant effort to normalize and integrate new operations. Today, Honey contains a large library of diverse operators that would be sufficient to solve most analytic tasks.
\end{compactenum}

Honey started as a simple tool developed for an experiment. Since then, it has grow and proved to be useful in many research tasks. Today Honey has allowed us, our colleagues, our students and our collaborators to quickly perform high quality complex and understandable analysis of a large variety of domains. Extrapolating from our experience, we hope and believe that Honey might change and improve the way we approach SSTS and other time series like data representations.
We also believe that Honey's core paradigm could be re-used in other domains and provide the same benefits (e.g. text processing, video processing, control, process definition).


\bibliographystyle{plain}
\bibliography{manuscript}

\end{document}